\documentclass[twocolumn,abs,prb,showpacs]{revtex4-1}

\usepackage{graphicx}
\usepackage{dcolumn}
\usepackage{float}
\usepackage{amsmath}
\usepackage{amssymb}
\usepackage{bm}
\usepackage[mathscr]{euscript}

\newcommand{\comment}[1]{}

\begin{document}


\title{Anomalous Behaviour in the Magneto-Optics of a Gapped Topological Insulator}

\author{C. J. Tabert$^{1,2}$}
\author{J. P. Carbotte$^{3,4}$}
\affiliation{$^1$Department of Physics, University of Guelph,
Guelph, Ontario N1G 2W1 Canada} 
\affiliation{$^2$Guelph-Waterloo Physics Institute, University of Guelph, Guelph, Ontario N1G 2W1 Canada}
\affiliation{$^3$Department of Physics, McMaster University,
Hamilton, Ontario L8S 4M1 Canada} 
\affiliation{$^4$Canadian Institute for Advanced Research, Toronto, Ontario M5G 1Z8 Canada}
\date{\today}

\begin{abstract}
{The Dirac fermions at the surface of a topological insulator can be gapped by introducing magnetic dopants.  Alternatively, in an ultra-thin slab with thickness on the order of the extent of the surface states, both the top and bottom surface states acquire a common gap value ($\Delta$) but with opposite sign.  In a topological insulator, the dominant piece of the Hamiltonian ($\hat{H}$) is of a relativistic nature.  A subdominant non-relativistic piece is also present and in an external magnetic field ($B$) applied perpendicular to the surface, the $N=0$ Landau level is no longer at zero energy but is shifted to positive energy by the Schr{\"o}dinger magnetic energy.  When a gap is present, it further shifts this level down by $-\Delta$ for positive $\Delta$ and up by $|\Delta|$ for a negative gap.  This has important consequences for the magneto-optical properties of such systems.  In particular, at charge neutrality, the lowest energy transition displays anomalous non-monotonic behaviour as a function of $B$ in both its position in energy and its optical spectral weight.  The gap can also have a profound impact on the spectral weight of the interband lines and on corresponding structures in the real part of the dynamical Hall conductivity.  Conversely, the interband background in zero field remains unchanged by the non-relativistic term in $\hat{H}$ (although its onset frequency is modified).
}
\end{abstract}


\maketitle

\section{Introduction}

The prediction and observation of two-dimensional (2D)\cite{Kane:2005,Kane:2005a,Bernevig:2006,Konig:2007} and three-dimensional (3D)\cite{Moore:2007,Fu:2007a,Hsieh:2008,Chen:2009} topological insulators (TIs) (and later topological crystalline insulators\cite{Fu:2011,Hsieh:2012,Tanaka:2012}) has lead to the discovery of many such materials\cite{Zhang:2009,Ando:2013}.  The subject remains of great interest because of its potential for the discovery of new physics\cite{Hasan:2010,Qi:2011,Moore:2010} and possible device applications such as in quantum computing\cite{Fu:2008}.  The dynamics of the helical surface states in a TI are dominated by a relativistic (Dirac) linear-in-momentum Hamiltonian which involves real spin as opposed to graphene\cite{Neto:2009} and other 2D membranes such as MoS$_2$\cite{Rose:2013,ZLi:2012} and buckled silicene\cite{Tabert:2013a} where it is the lattice pseudospin which is involved.  While the Dirac cones in graphene display perfect particle-hole symmetry, they do not in a TI due to the presence of an additional non-relativistic contribution.  This term is quadratic in momentum and is described by a Schr\"odinger mass $m$.  Typically, in Bi$_2$Se$_3$ and Bi$_2$Te$_3$ as examples, $m=0.16m_e$ and $0.09m_e$\cite{Liu:2010,ZLi:2013}, respectively, where $m_e$ is the bare electron mass.  The relativistic contribution to the total Hamiltonian has its origin in a strong spin-orbit coupling and leads to the phenomenon of spin-momentum locking with the inplane spin component perpendicular to the direction of the momentum\cite{Hsieh:2009,Nishide:2010}.  The subdominant Schr\"odinger contribution distorts the perfect Dirac cone into an hourglass shape with the cross-section of the upper conduction-band cone decreasing with increasing energy while that of the valence band increases with decreasing energy below the Dirac point.  The spin-momentum locking is retained.

It is possible to introduce a gap ($\Delta$) at the Dirac point of the surface fermion band structure.  This can be achieved by the introduction of magnetic dopants.  In the work of Chen \emph{et al.}\cite{Chen:2010}, a gap of $\sim 7$ meV was seen in (Bi$_{0.99}$Mn$_{0.01}$)$_2$Se$_3$ using angular-resolved photoemmision spectroscopy (ARPES).  A significantly larger gap can be generated in an ultrathin TI slab when its thickness is sufficiently reduced such that the top and bottom surface states hybridize\cite{Lu:2010,Linder:2009,Shan:2010}.  In Ref.~\cite{Lu:2010}, it is shown that the sign of the induced gap can change from a negative value to a positive one as the distance between the two surfaces is reduced below $\sim 25$ \AA.  Additionally, the magnitude can grow to $\sim 100$ meV.  The Hamiltonians for describing the top and bottom surfaces in a thin slab are related by the mapping $\Delta\rightarrow-\Delta$\cite{Lu:2010}.  A gap of superconducting origin can also be created in a TI through proximity to a superconductor\cite{Wang:2013} but this goes beyond the scope of the present manuscript.  When a TI has a finite gap, a $z$-component of spin perpendicular to its surface is generated.  This implies that the inplane spin component involved in the locked spin texture is reduced.  This effect has recently been observed in spin-polarized ARPES data\cite{Neupane:2014}.  

The quantum Hall effect of an ultrathin TI has been discussed by Yoshimi \emph{et al.}\cite{Yoshimi:2015} and Zhang \emph{et al.}\cite{Zhang:2015}.  Oscillations in the quantum capacitance of such systems feature in the work of Tahir \emph{et al.}\cite{Tahir:2013b} and thermoelectric transport in the work of Tahir and Vasilopoulos\cite{Tahir:2015}.  Other studies include work on the Kerr and Faraday effects in thin films with broken time-reversal symmetry\cite{Tse:2010a,Tse:2010} and magneto-optical transport\cite{Tkachov:2011,Tkachov:2011a}.

The dynamical conductivity can provide detailed information on electron dynamics in 2D systems.  An example is graphene where experiments have verified the predicted universal background\cite{Li:2008,Nair:2008} as well as revealed details about correlations due to electron-phonon interactions\cite{Carbotte:2010,Stauber:2008}.  The magneto-optics of such systems provide additional information\cite{ZLi:2013,Tabert:2015b,Sadowski:2007,Jiang:2007,Deacon:2007,Gusynin:2007,Gusynin:2007b,Orlita:2010,Schafgans:2012}.  In this paper, we consider the magneto-optics of a gapped TI.  Both the longitudinal and transverse Hall conductivity are considered.

The paper is structured as follows.  In Sec.~II, we specify our Hamiltonian which includes a relativistic and non-relativistic kinetic energy term as well as a gap.  On a given surface, for simplicity, we treat a single Dirac cone centred at the $\Gamma$ point of the surface Brillouin zone.  This applies to Bi$_2$Se$_3$, Bi$_2$Te$_3$ and other similar systems; however, for materials such as samarium hexaboride\cite{Roy:2014}, three such cones are present.  To treat an ultrathin slab with a tunnelling gap between top and bottom surfaces, we need to consider both positive and negative gap values\cite{Lu:2010}.  We solve for the eigenvalues of the associated Landau levels (LLs) which emerge when a perpendicular magnetic field ($B$) is applied to the surface of a TI.  The corresponding eigenvectors are also reported.  Particular attention is given to the $N=0$ LL which behaves quite differently in a TI than it does for graphene.  In particular, even without a gap, the LL no longer exhibits particle-hole symmetry.  The $N=0$ level has moved to positive energy from its zero-energy value in the pure relativistic case.  It is important to realize that two competing magnetic energy scales exist in a TI.  The dominant magnetic energy scale is that associated with the relativistic term ($E_1=\hbar v_F\sqrt{eB/\hbar}$).  The second is $E_0=\hbar eB/m$ and comes from the Schr\"odinger mass.  Here $v_F$ is the Dirac Fermi velocity and $m$ is the Schr\"odinger mass.  It is clear that as $B$ increases, the relative importance of $E_0$ increases.  However, at one Tesla, it is an order of magnitude smaller than $E_1$.  Nevertheless, as we will highlight, $E_0$ introduces important changes to some aspects of the magneto-optics while others are left unchanged.  In Sec.~III, we provide the results of a Kubo formula approach to the dynamical conductivity.  We give explicit expressions for Re$\sigma_{xx}(\Omega)$ and Re$\sigma_{xy}(\Omega)$ at finite $B$ and discuss the modifications needed to obtain the respective imaginary terms.  Numerical results are presented when a finite gap is included and are compared with the pure relativistic limit. Emphasis is placed on the real part of the Hall conductivity.  Both charge neutrality and finite chemical potential are described as is the effect of changing the value of $\Delta$ and $B$.  In Sec.~IV, particular attention is given to the optical spectral weight of the various magneto-optical absorption lines and how the gap impacts these values.  For a fixed gap, it is found that the position of the intraband line and its spectral weight show a jump at a critical value of magnetic field $B_c=(2\Delta m)/(\hbar e[1+g])$ where $g$ is a Zeeman splitting contribution.  This jump only occurs when the sign of the gap is positive.  In Sec.~V, we derive a simplified formula for the optical spectral weight of the intraband transitions when $B\rightarrow 0$ and the chemical potentials $\mu$ is much greater than both magnetic energy scales.  This is obtained from our general conductivity formulas for finite $B$.  In Sec.~VI, we turn to a discussion of the interband transitions and how they evolve into a universal background which is independent of the Schr\"odinger mass.  We discuss how the gap modifies this result.  A summary and conclusions follow in Sec.~VII.

\section{Formalism}

\subsection{No Magnetic Field}

In the absence of a magnetic field, the model Hamiltonian for describing the helical surface states of a TI is given by\cite{Bychkov:1984,Bychkov:1984a}
\begin{align}\label{HAM}
\hat{H}=\frac{\hbar^2 k^2}{2m}+\hbar v_F(k_x\hat{\sigma}_y-k_y\hat{\sigma}_x)+\Delta\hat{\sigma}_z,
\end{align}
where $\hat{\sigma}_x,\hat{\sigma}_y,\hat{\sigma}_z$ are the Pauli-spin matrices and $\hbar\bm{k}$ is the momentum relative to the $\Gamma$ point of the surface Brillouin zone.  The first term is a quadratic-in-momentum non-relativistic kinetic energy piece with electronic mass $m$.  The second is relativistic and linear-in-momentum with Fermi velocity $v_F$.  The last term (which provides a gap) is present when the surface of the TI is doped with magnetic particles or, alternatively, when the TI is made ultrathin so that the top and bottom surface states hybridize.  In the latter case, the two surfaces have a gap of the same magnitude but opposite sign\cite{Lu:2010}.  Appropriate parameters for the Hamiltonian have been provided by band structure calculations and, as an illustration, $mv_F^2=96$ meV (130 meV) with $m=0.09m_e$ ($0.16m_e$) in Bi$_2$Te$_3$ (Bi$_2$Se$_3$).  From these parameters, one can compute other useful parameters such as the Dirac and Schr\"odinger magnetic energies $E_1\equiv\hbar v_F\sqrt{eB/\hbar}$ and $E_0=\hbar eB/m$, respectively.  They are 10.9 meV (12.7 meV) and 1.25 meV (0.7 meV) for Bi$_2$Te$_3$ (Bi$_2$Se$_3$) for $B=1$T.

Through a $1\%$ replacement of Bi with Mn on the surface of Bi$_2$Se$_3$, Chen \emph{et al.}\cite{Chen:2010} produced a gap of $\Delta\sim 7$ meV.  Much larger gap values can be obtained by hybridization of the top and bottom surface states in an ultrathin slab as estimated in the work of Lu \emph{et al.}\cite{Lu:2010}.  For their parameter set, the Fermi velocity is of order $6.2\times 10^5$ m/s with some variation on slab thickness ($L$).  They find $\Delta$ can be of order 50 meV and can even change sign ($\Delta<0$ for $L\sim 32$ \AA\, and $\Delta>0$ below 25 \AA).

\subsection{Finite Magnetic Field}

To account for the influence of an external magnetic field $B$ applied perpendicular to the surface of a TI, we employ the Landau gauge with vector potential $\bm{A}=(0,Bx,0)$.  Including a Zeeman term to the Hamiltonian of the form $(-1/2)g_s\mu_BB\hat{\sigma}_z$, with $g_s$ the Zeeman splitting strength ($g_s\sim 8$ for Bi$_2$Se$_3$\cite{Wang:2010}) and $\mu_B=e\hbar/(2m_e)$ the Bohr magneton.  The LL energies are
\begin{align}\label{LL-TI}
\mathcal{E}_{N,s}=\left\lbrace\begin{array}{cc}
\displaystyle E_0N+s\sqrt{2NE_1^2+\left[\Delta-\frac{E_0}{2}(1+g)\right]^2} & N=1,2,3,...\\
\displaystyle\frac{E_0}{2}(1+g)-\Delta & N=0
\end{array}\right.,
\end{align}
where $s=\pm$ for the conduction and valence band, respectively, and $g=g_sm/(2m_e)$ is the renormalized Zeeman-coupling coefficient.  The relativistic and non-relativistic magnetic energy scales are $E_1=\hbar v_F\sqrt{eB/\hbar}$ and $E_0=\hbar eB/m$, respectively.  The associated eigenvectors are
\begin{align}\label{wavefunc}
\left|Ns\right\rangle=\left(\begin{array}{c}
\mathcal{C}^\uparrow_{N,s}\left|N-1\right\rangle\\
\mathcal{C}^\downarrow_{N,s}\left|N\right\rangle
\end{array}\right),
\end{align}
where
\begin{align}\label{Cup}
\mathcal{C}^\uparrow_{N,s}=\left\lbrace\begin{array}{cc}
\displaystyle -s\sqrt{\frac{1}{2}+s\frac{\Delta-(1+g)E_0/2}{2(\mathcal{E}_{N,+}-E_0N)}} & N=1,2,3,...\\
0 & N=0
\end{array}\right.,
\end{align}
and\begin{align}\label{Cdown}
\mathcal{C}^\downarrow_{N,s}=\left\lbrace\begin{array}{cc}
\displaystyle\sqrt{\frac{1}{2}-s\frac{\Delta-(1+g)E_0/2}{2(\mathcal{E}_{N,+}-E_0N)}} & N=1,2,3,...\\
1 & N=0
\end{array}\right..
\end{align}
We begin our analysis by emphasizing that, for gapless graphene (no Schr\"odinger term in Eqn.~\eqref{HAM} and $\Delta=0$), the $N=0$ LL falls at $\mathcal{E}=0$ in the absence of Zeeman splitting. For a TI, however, the non-relativistic term in the Hamiltonian pushes this level to positive energy as does the inclusion of the Zeeman term.  By contrast, including a gap pushes the $N=0$ level down in energy for $\Delta>0$ and up for $\Delta<0$.  This plays an important role in the following considerations.

\section{Magneto-Optical Conductivity}

\begin{figure}[h!]
\begin{center}
\includegraphics[width=0.9\linewidth]{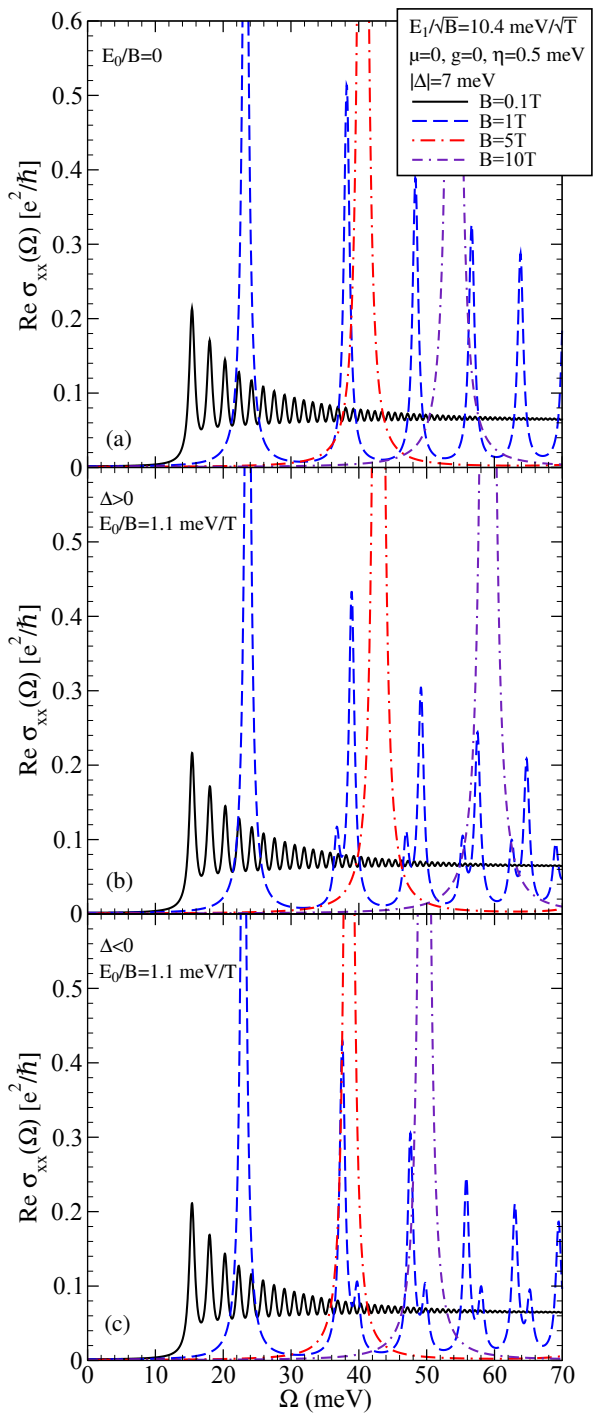}
\end{center}
\caption{\label{fig:Condxx-Dpm7}(Color online) Longitudinal magneto-optical conductivity of a (a) gapped Dirac system ($E_0=0$) compared with that of a TI for a (b) positive  and (c) negative gap. Various values of $B$ are considered; for small $B$, the familiar interband background emerges.
}
\end{figure}

Based on the Kubo formula in the one-loop approximation, the longitudinal ac conductivity $\sigma_{xx}(\Omega)$ takes the form\cite{Tabert:2015b}
\begin{align}\label{sigmaxx}
{\rm Re}\left\lbrace\frac{\sigma_{xx}(\Omega)}{e^2/\hbar}\right\rbrace &=\frac{E_1^2}{2\pi}\sum_{\substack{N,M=0 \\s,s^\prime=\pm}}^\infty\frac{f_{M,s^\prime}-f_{N,s}}{\mathcal{E}_{N,s}-\mathcal{E}_{M,s^\prime}}\\
&\times\frac{\eta}{(\Omega+\mathcal{E}_{M,s^\prime}-\mathcal{E}_{N,s})^2+\eta^2}\notag\\
&\times[\mathcal{F}(Ns;Ms^\prime)\delta_{N,M-1}+\mathcal{F}(Ms^\prime;Ns)\delta_{M,N-1}],\notag
\end{align}
where $\eta\equiv\hbar/(2\tau)$ is a phenomenological optical scattering rate and the optical matrix element is
\begin{align}\label{mat-el}
\mathcal{F}(Ns;Ms^\prime)&\equiv\left[\mathcal{C}^\uparrow_{M,s^\prime}\mathcal{C}^\downarrow_{N,s}-\frac{E_0}{\sqrt{2}E_1}\right.\\
&\times\left.\left(\sqrt{N}\mathcal{C}^\uparrow_{M,s^\prime}\mathcal{C}^\uparrow_{N,s}+\sqrt{N+1}\mathcal{C}^\downarrow_{M,s^\prime}\mathcal{C}^\downarrow_{N,s}\right)\right]^2.\notag
\end{align}
For the imaginary part of the longitudinal conductivity, the $\eta$ in the numerator of Eqn.~\eqref{sigmaxx} is replaced by $\Omega+\mathcal{E}_{M,s^\prime}-\mathcal{E}_{N,s}$.  The Hall conductivity also follows with small modifications to Eqn.~\eqref{sigmaxx}.  Its real part is given by Eqn.~\eqref{sigmaxx} with the $\eta$ in the numerator replaced by $\Omega+\mathcal{E}_{M,s^\prime}-\mathcal{E}_{N,s}$ and with a switch in sign between the matrix elements.  For the imaginary part, the numerator is $-\eta$ rather than the energy difference plus photon energy.  In Eqn.~\eqref{sigmaxx}, $f_{N,s}$ is the thermal occupation factor which reduces to the Heaviside step function $\Theta(\mu-\mathcal{E}_{N,s})$ at zero-temperature, where $\mu$ is the chemical potential.  Our numerical results for the real part of the longitudinal conductivity are presented in Fig.~\ref{fig:Condxx-Dpm7} for $\mu=0$, $g=0$, and $|\Delta|=7$ meV.  
In the numerics, we use a broadening parameter $\eta=0.5$ meV.  The magnetic field values used are 0.1T (solid black curve), 1T (dashed blue), 5T (dash-dotted red), and 10T (double-dash-dotted purple). Frame (a) is for comparison and has $E_0=0$ (i.e. no Schr\"odinger contribution so that we would be dealing with the particle-hole-symmetric Dirac case when the gap is neglected).  When the magnetic field is small, the distance in energy between LLs (and consequently between optical absorption lines) is small.  This arises from the $\sqrt{B}$ dependence of the dominant magnetic energy scale $E_1$.  At higher photon energies,the well known universal background provided by interband transitions is revealed in the solid black curve.  In our units, its height is $1/16$ which (when multiplied by a factor of $4$ for spin and valley degeneracy) gives the expected graphene value.  As the photon frequency is reduced, the amplitude of the oscillations about the background value increase as does the energy spacing.  The energy of the first absorption line is at $\Omega=\Delta+\sqrt{2E_1^2+\Delta^2}$ and, for $B=0.1$T, is already close to its $B=0$ limiting value of $2\Delta$.  We also note that the average interband background grows from its universal value of $1/16$ to twice this amount at the gap edge.  This is somewhat obscure in the figure as the quantum oscillations are rather large in this energy range.  In graphene, it is known that the introduction of a gap modifies that universal background by a multiplicative factor of $(\Omega^2+4\Delta^2)/\Omega^2$.  Here, we find that this holds for a TI even when a subdominant Schr\"odinger term is included in the Hamiltonian.  As the magnitude of the magnetic field is increased, the first absorption line moves to higher photon frequency.  For example, in the double-dash-dotted purple curve, it has moved to $\sim 54$ meV.  

For the pure relativistic case, the sign of the gap makes no difference in Re$\sigma_{xx}(\Omega)$.  This changes in a TI as can be seen in frames (b) and (c).  It is particularly striking that the position of the first peak in the double-dash-dotted purple curve has moved to higher energy for $\Delta>0$ and to lower energies for $\Delta<0$ relative to the pure Dirac system even though all other parameters are kept the same.  The only difference is that now the Schr\"odinger magnetic energy scale $E_0$ is nonzero.  Even more striking is the fact that, except for the first process, all other absorption lines show a satellite peak attached to each dominant peak.  In frame (b) ($\Delta>0$), this satellite line is below the main absorption process of the doublet.  Conversely, in frame (c), the relative locations are reversed.  These subdominant peaks are a clear signature of the small non-relativistic term in the Hamiltonian.  They have already been noted in the previous work of Li and Carbotte\cite{ZLi:2013} when $\Delta=0$. In that case, however, the optical spectral weight in each of the two lines is nearly equal.  The opening of a band gap substantially changes this and will be studied in detail in the following section.  While both frames (b) and (c) include the effect of a finite Schr\"odinger mass, the universal limit is unchanged for $\Omega\gg|\Delta|$ and is also unaffected by the sign of the gap.  The introduction of $E_0$ only slightly affects the conductivity for small $B$.  This is because the factor $\Delta-(1+g)E_0/2$ appearing in the LL energies is nearly unchanged from its pure relativistic value of $\Delta$ (for $g=0$).  For example, when $B=0.1$T (solid black), $E_0/2\approx 0.055$ meV which is much smaller than the damping $\eta=0.5$ meV used in our numerics.

\begin{figure}[h!]
\begin{center}
\includegraphics[width=0.9\linewidth]{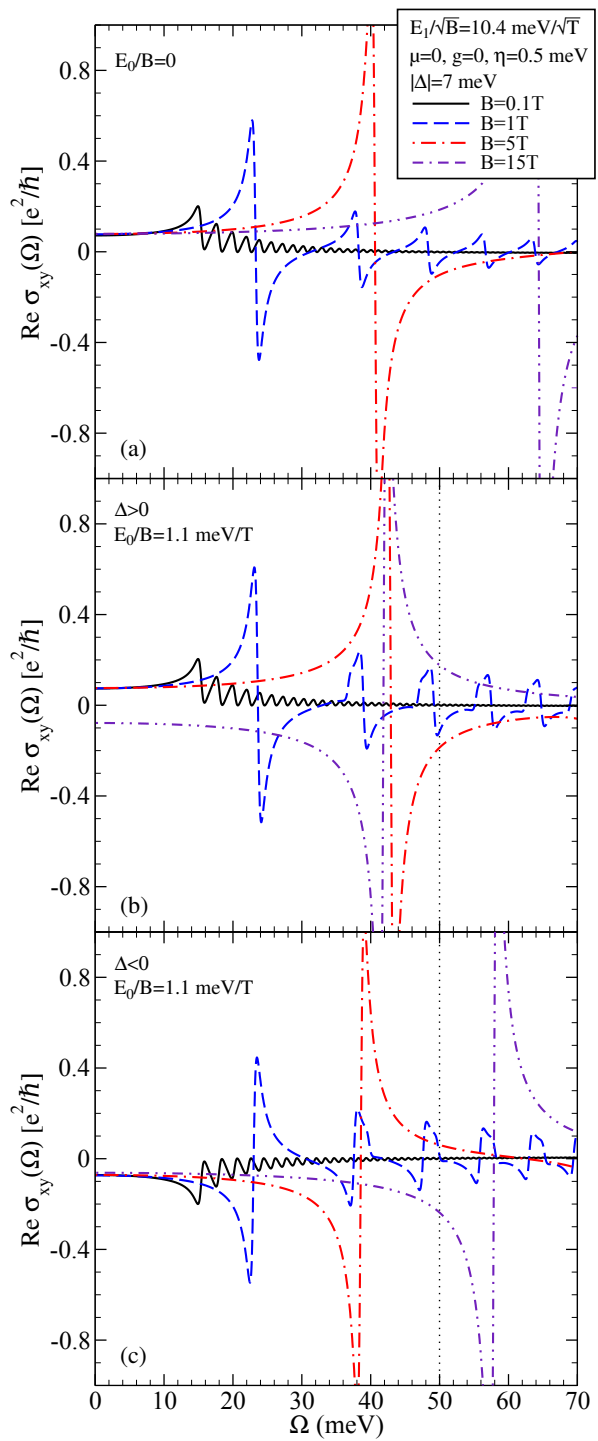}
\end{center}
\caption{\label{fig:Condxy-Dpm7}(Color online) Transverse magneto-optical conductivity of a (a) gapped Dirac system ($E_0=0$) compared with that of a TI for a (b) positive and (c) negative gap. Various values of $B$ are considered.  The $\Omega=0$ limit gives the quantized Hall plateaus.  For $\Delta>0$, the sign of the dc Hall effect in a TI switches as $B$ becomes large.
}
\end{figure}

In Fig.~\ref{fig:Condxy-Dpm7}, we show results for the real part of the Hall conductivity Re$\sigma_{xy}(\Omega)$ for the same parameters as Re$\sigma_{xx}(\Omega)$ except instead of $B=10$T we employ 15T to accentuate the important features.
Frame (a) shows the pure relativistic result while frames (b) and (c) display finite $E_0$ and $\Delta=7$ meV and $-7$ meV, respectively.  As in Fig.~\ref{fig:Condxx-Dpm7}, the solid black curve shows small oscillations which become somewhat more pronounced as $\Omega$ is reduced but, nevertheless, define the background fairly well.  The $B=0$ background is basically zero for large photon energy and has a characteristic peak-valley structure at small $\Omega$.  We wish to emphasize several other features.  First, the dc limiting value of Re$\sigma_{xy}(\Omega)$ gives the Hall plateaus which are quantized.  As discussed in Ref.~\cite{ZLi:2014}, in the context of the conductivity and in Refs.~\cite{Tabert:2015,Tabert:2015b} in the context of the magnetization, the introduction of a subdominant non-relativistic term does not change this quantization. It keeps its relativistic value independent of $m$.  In our units, this is $1/2$ times $1/(2\pi)$ at charge neutrality which agrees with what is known for graphene when a valley-spin-degeneracy factor of $4$ is included.  This value is independent of $B$ but can be made to change sign as $B$ is increased in the case when $\Delta>0$ as can be seen in frame (b) for $B=15$T (dash-double-dotted purple).  It has switched from 1/(4$\pi$)$e^2/\hbar$ for $B=0.1$, 1 and 5T to the negative of the same value.  This is to be contrasted with the $\Delta<0$ case where it is always equal to $-1/(4\pi)e^2/\hbar$.  This difference in sign implies additional differences in the curves at finite photon energies.  For $\Delta=0$, the first LL is positioned at $(1+g)E_0/2$ at which energy, the Hall conductivity would jump from $-1/2$ to $1/2$ (in units of $e^2/h$).  For nonzero $\Delta$, this transition energy is instead moved to $-\Delta+(1+g)E_0/2$ (i.e. down for $\Delta>0$ and up for $\Delta<0$).  This means that in frame (b), we have Re$\sigma_{xy}(\Omega=0)=(1/2)e^2/h$ for fields less than the critical value $B_c=2\Delta m/([1+g]\hbar e)\sim 13$T and $-(1/2)e^2/h$ above the critical field (dash-double dotted purple).  By contrast, for negative gap values, $-\Delta+(1+g)E_0/2$ remains positive so that the Hall quantization retains its value of $-(1/2)e^2/h$ for all $B$. In the pure relativistic case, the sign of the Hall conductivity cannot be changed by increasing $B$ for $g=0$.  This sign change is a signature of the subdominant magnetic energy scale $E_0$ in the TI Hamiltonian or of a Zeeman term.  By choice, we have taken $g=0$ in all the curves shown in Fig.~\ref{fig:Condxy-Dpm7}.  There are other features in this figure which require comments.  First, the dashed blue curve starts with a peak-valley structure when the Hall quantization is $e^2/(2h)$ [frames (a) and (b)] while in the lower frame, when Re$\sigma_{xy}(\Omega)=-e^2/(4\pi\hbar)$, this structure is inverted.  The peak-valley feature is followed by a series of other similar structures as the photon energy is increased.  These have a different shape in a TI (lower two frames) compared to the pure Dirac system (top frame).  Each subsequent peak is not as sharp as in the frame (a) but shows a rather flat top; this can be traced to the splitting of a single line into doublets (see the lower frames of Fig.~\ref{fig:Condxx-Dpm7}).  Also, for the TI, there is a clearly defined knee just before (above) a new peak sets in when $\Delta>0$ ($\Delta<0$).  Using the vertical dotted line at $\Omega=50$ meV as a guide, we see that the peak in the middle frame is followed by a sharp drop and a minimum while, in the lower frame, there is knee and the minimum following at higher photon frequencies.

\begin{figure}[h!]
\begin{center}
\includegraphics[width=1.0\linewidth]{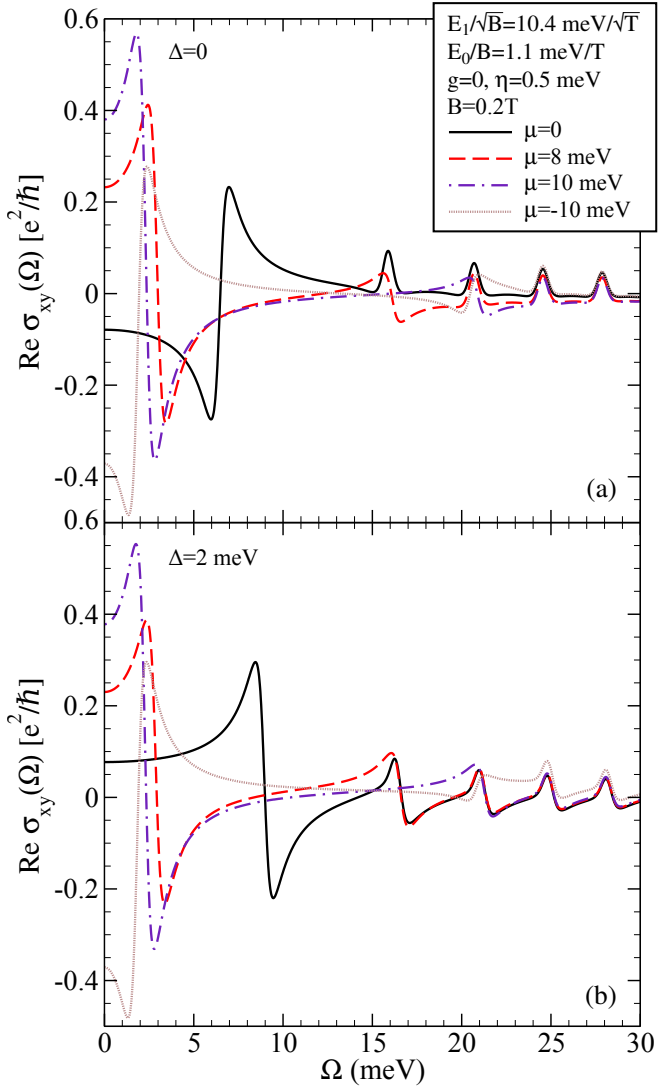}
\end{center}
\caption{\label{fig:Condxx-xy-D2}(Color online) Real part of the transverse Hall conductivity of a TI for (a) $\Delta=0$ and (b) $\Delta=2$ meV for various $\mu$. The dc Hall effect steps through different quantized values as $\mu$ changes.
}
\end{figure}
In Fig.~\ref{fig:Condxx-xy-D2}, we show the real part of the Hall conductivity for various values of chemical potential.  Again, the dc limit gives the quantization of the Hall plateaus.  Comparing $\Delta=0$ [Fig.~\ref{fig:Condxx-xy-D2}(a)] with $\Delta=2$ meV [Fig.~\ref{fig:Condxx-xy-D2}(b)], we see that this intercept is robust and does not change with the value of $\Delta$.  Instead, it retains the value associated with graphene (up to a degeneracy factor of 4) even in the presence of the non-relativistic Schr\"odinger mass term. However, the presence of a small gap can change the sign of the dc Hall effect from $-e^2/(2h)$ in Fig.~\ref{fig:Condxx-xy-D2}(a) to $e^2/(2h)$ in Fig.~\ref{fig:Condxx-xy-D2}(b) for $\mu=0$ (solid black curve).  The first peak-valley structure also flips to a valley-peak feature.  Its location in energy has shifted from 7 meV to 9 meV because of the gap.  We wish to stress that, as $\mu$ is increased, the quantization of the Hall conductivity (in units of $e^2/h$) increases from 1/2 to 3/2 to 5/2 as another LL is crossed.  Here, for $\mu=8$ meV (10 meV), the 3/2 (5/2) plateau is involved [dashed red (dash-dotted purple) curve].  For $\mu=-10$ meV (dotted brown curve), the Hall intercept is $-5/2$ and instead of a peak-valley structure following it is a valley-peak feature.  Higher energy structures are also modified by the gap.

\section{Optical Spectral Weight}

\begin{figure}[h!]
\begin{center}
\includegraphics[width=1.0\linewidth]{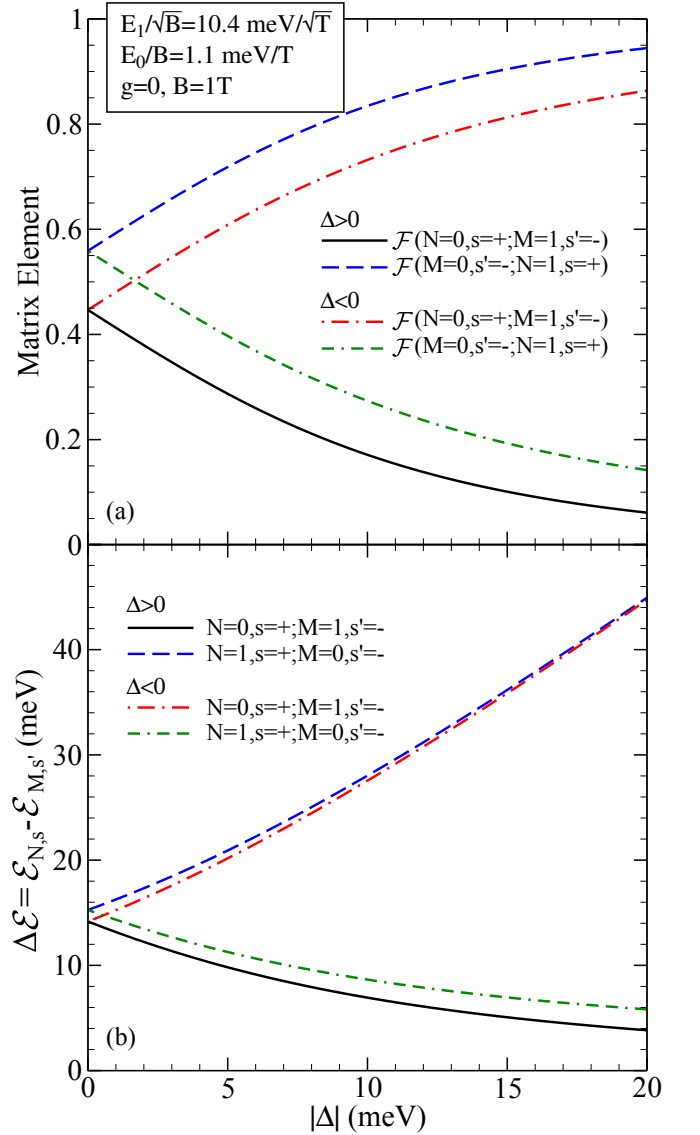}
\end{center}
\caption{\label{fig:Mat-El-01}(Color online) (a) Optical matrix element and (b) transition energy of a TI as a function of $|\Delta|$ for positive and negative gaps.  The results for the $0\rightarrow 1^+$ transition are compared to $1^-\rightarrow 0$ (clear asymmetry is observed).
}
\end{figure}

In this section, we return to the question of the optical spectral weight under the absorption peaks of Re$\sigma_{xx}(\Omega)$ and how it is distributed amongst the two peaks of the doublet.  To understand these features, it is useful to look at the optical matrix element $\mathcal{F}(Ns;Ms^\prime)$ of Eqn.~\eqref{mat-el} which, along with the energy difference $\mathcal{E}_{N,s}-\mathcal{E}_{M,s^\prime}$, gives the optical spectral weight associated with the real part of the longitudinal conductivity [Eqn.~\eqref{sigmaxx}].  This is shown in Fig.~\ref{fig:Mat-El-01} for the intraband line.  
In frame (a), we plot $\mathcal{F}(N=0,s=+;M=1,s^\prime=-)$ which applies to the transition $1^-\rightarrow 0$ and $\mathcal{F}(M=0,s^\prime=-;N=1,s=+)$ which corresponds to $0\rightarrow 1^+$.  There are two cases of interest.  For $\Delta>0$, the solid black curve gives the $1^-\rightarrow 0$ transition and is seen to decrease with increasing $\Delta$ while the dashed blue curve is for $0\rightarrow 1^+$ and increases with $\Delta$.  Even at $\Delta=0$, it is above the solid black line.  A similar pair of curves describes the optical matrix elements for $\Delta<0$.  These are the dashed-dotted red and double-dash-dotted green curves for $1^-\rightarrow 0$ and $0\rightarrow 1^+$, respectively.  However, now the former increases with $\Delta$ while the latter decreases.  This causes them to cross.  The energy difference between the two LLs involved in a given transition ($\Delta\mathcal{E}=\mathcal{E}_{N,s}-\mathcal{E}_{M,s^\prime}$) is also of interest for two reasons.  First, it gives information on the required photon energy to excite this transition in absorptive experiments.  Secondly, it is the ratio of $\mathcal{F}$ to $\Delta\mathcal{E}$ which determines the optical spectral weight of the line.  The thermal factors in Eqn.~\eqref{sigmaxx} at $T=0$ are either zero or one while the Lorentzian factor gives $\pi$ once integrated to get the optical spectral weight.   The remaining $E_1^2$ is constant at fixed $B$.  The energy difference is plotted in Fig.~\ref{fig:Mat-El-01}(b) using the same line types as in the top frame.  Similar to $\mathcal{F}$, for $\Delta>0$, the energy of the $1^-\rightarrow 0$ transition (solid black) decreases with increasing $\Delta$ while it increases for the $0\rightarrow 1^+$ transition (dashed blue).  For negative gaps, $\Delta\mathcal{E}$ also shows the same trend as for $\mathcal{F}$.  Importantly, the spectral weight factor $\mathcal{F}/\Delta\mathcal{E}$ decreases with $|\Delta|$ in all cases.  This identifies the factors which contribute to the intensity decay of the intraband line with increasing $\Delta$.  This also explains the increase in photon energy needed to excite this transition.

\begin{figure}[h!]
\begin{center}
\includegraphics[width=1.0\linewidth]{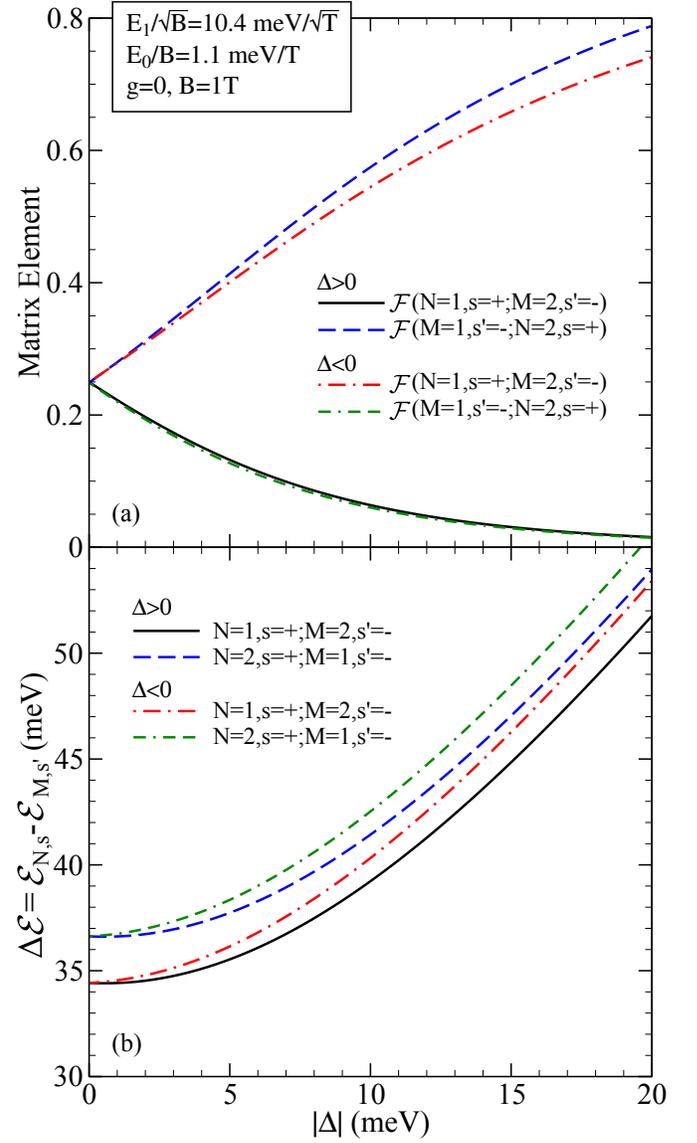}
\end{center}
\caption{\label{fig:Mat-El-12}(Color online) (a) Optical matrix element and (b) transition energy of a TI as a function of $|\Delta|$ for positive and negative gaps.  The results for the $1^-\rightarrow 2^+$ transition are compared to $2^-\rightarrow 1^+$ (asymmetry is observed).
}
\end{figure}

Another interesting feature noted in reference to Fig.~\ref{fig:Condxx-Dpm7} for Re$\sigma_{xx}(\Omega)$ for several $B$ was the splitting of the interband lines into doublets with the intensity of the lines being distinctly different (the higher $\Omega$ line being larger for $\Delta>0$).  Similarly, the lower energy line of the doublet loses intensity as $\Delta>0$ is increased.  As previously mentioned, it is the ratio of the matrix elements $\mathcal{F}$ and energy denominators $\Delta\mathcal{E}$ which determines the optical intensity (up to a factor) of the spectral lines.  In Fig.~\ref{fig:Mat-El-12}, we plot $\mathcal{F}$ [frame (a)] and $\Delta\mathcal{E}$ [frame (b)] for the $2^-\rightarrow 1^+$ and $1^-\rightarrow 2^+$ transitions.  
For $\Delta>0$, the $2^-\rightarrow 1^+$ optical matrix element decreases with increasing $\Delta$ while it increases for $1^-\rightarrow 2^+$.  The opposite behaviour is found for $\Delta<0$ as expected.  Concerning the energies of the lines, we see in the lower frame that they all increase with increasing $|\Delta|$.  In both cases, the $2^-\rightarrow 1^+$ transition is the lower energy line of the doublet while the $1^-\rightarrow 2^+$ line has higher energy.  The energy associated with the line splitting is reasonably constant at $|\Delta|$.  These facts conform with what we have found in other figures.  For small $|\Delta|$, the intensity of both the $2^-\rightarrow 1^+$ and $1^-\rightarrow 2^+$ lines is almost the same.  However, as $\Delta>0$ is increased the energy of the $2^-\rightarrow 1^+$ (lower peak of the doublet) rapidly decreases while that of the $1^-\rightarrow 2^+$ line (upper peak in the pair) increases.  It is opposite for $\Delta<0$.  This rapid change is attributed to a sharp increase/decrease of the optical matrix element $\mathcal{F}$ which depends on the sign of the gap and the transition involved [as seen in Fig.~\ref{fig:Mat-El-12}(a)].

\begin{figure}[h!]
\begin{center}
\includegraphics[width=1.0\linewidth]{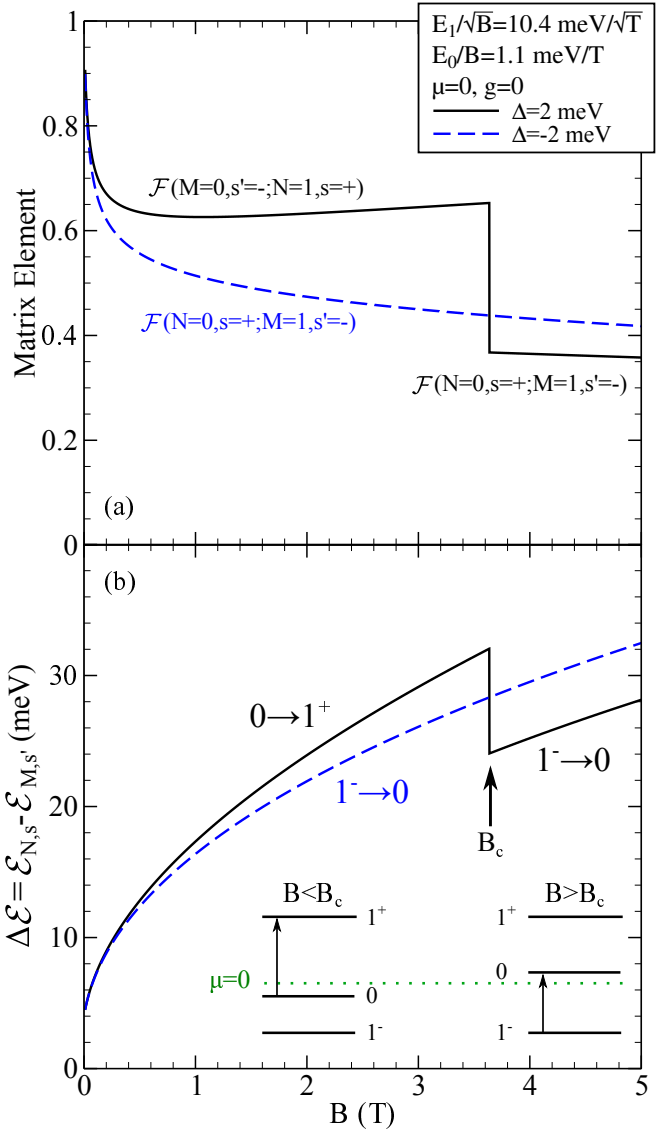}
\end{center}
\caption{\label{fig:Ediff-D2}(Color online) (a) Optical matrix element and (b) transition energy of a TI as a function of $B$ for positive and negative gaps.  A discontinuity is present for $\Delta>0$ at $B=B_c$ due to a change in sign of the $N=0$ LL energy [inset of (b)].  This changes the charge neutral transition from $0\rightarrow 1^+$ to $1^-\rightarrow 0$. 
}
\end{figure}

While it has been useful to consider variations of $\Delta$ in Figs.~\ref{fig:Mat-El-01} and ~\ref{fig:Mat-El-12}, an anomalous behaviour of the energy and intensity of the intraband line at charge neutrality is best brought out by considering a fixed value of $\Delta$ and varying magnetic field.  This is shown in Fig.~\ref{fig:Ediff-D2} for $g=0$, $\mu=0$ and $\Delta=\pm 2$ meV.
Here, the occupation factors $f_{Ms^\prime}$ and $f_{Ns}$ in Eqn.~\eqref{sigmaxx} play an essential role.  At zero temperature, these are just Heaviside step functions which equal 1 if the state is occupied and 0 if it is empty.  In the inset of Fig.~\ref{fig:Ediff-D2}(b), we show the lowest LLs involved as well as the location of zero energy (dotted line) where $\mu$ falls by arrangement.  For $\Delta>0$, we wish to emphasize that the zeroth level falls at an energy $(1+g)E_0/2-\Delta$.  For $\Delta=0$, this level has a positive energy while for $\Delta>0$ it can be moved to negative energies by decreasing the external magnetic field below $B_c=2\Delta m/([1+g]\hbar e)$ as previously noted.  For $B<B_c$, the optical transition involved in the intraband response is $0\rightarrow 1^+$.  For $B>B_c$, this switches to $1^-\rightarrow 0$.  These two transitions have different intensities and correspond to different photon energies as seen in the solid black curve in Fig.~\ref{fig:Ediff-D2}(a) for the matrix element $\mathcal{F}$ and in frame (b) for the energy $\Delta\mathcal{E}$.  Both of these quantities have a discontinuity at $B=B_c$.  The intensity and energy of the line drop above the critical field.  This effect is only present for $\Delta>0$.  For the negative gap case (dashed blue curve), there is no jump at any value of $B$.  Note that this behaviour directly depends on the presence of a subdominant non-relativistic magnetic energy $E_0$.  This anomalous behaviour is one of our important results.

\section{The Limit of Small $B$}

We now consider the limit of $B\rightarrow 0$ with a view of understanding the effects introduced by a gap.  We consider both the intraband transitions which give the Drude peak when $B=0$ and the interband transitions which provide a background.  We begin with the intraband response of the longitudinal conductivity [see Eqn.~\eqref{sigmaxx}].  For simplicity, we take $\mu>0$ and assume both the relativistic and non-relativistic magnetic energy scales to be small in comparison to $\mu$.  After employing the Kronecker $\delta$-functions, only the sum over $N$ remains.  We define a critical value of $N$ (denoted $N_c$) such that $\mathcal{E}_{N_c,+}=\mu$.  We also integrate over $\Omega$ from 0 to $\infty$ to get the total optical spectral weight under the intraband line:
\begin{align}
W_D=\frac{e^2}{\hbar}\frac{E_1^2}{2}\sum_{N}\left(-\frac{\partial f_{N,+}}{\partial\mathcal{E}_{N,+}}\right)\mathcal{F}(N-1,+;N,+).
\end{align}
At zero temperature, the derivative of the Fermi-Dirac function becomes a Dirac $\delta$-function $\delta(\mathcal{E}_{N,+}-\mu)$ which we use to carry out the sum over $N$ (which becomes an integral in the $B\rightarrow 0$ limit).  That is,
\begin{align}
W_D&=\frac{e^2}{\hbar}\frac{E_1^2}{2}\int_0^\infty dN\left(\frac{\partial\mathcal{E}_{N,+}}{\partial N}\right)^{-1}\delta(\mu-\mathcal{E}_{N,+})\notag\\
&\quad\quad\quad\quad\quad\times\mathcal{F}(N_c-1,+;N_c,+)\notag\\
&=\frac{e^2}{\hbar}\frac{E_1^2}{2}\left(\frac{\partial\mathcal{E}_{N,+}}{\partial N}\right)^{-1}\bigg|_{N_c}\mathcal{F}(N_c-1,+;N_c,+).
\end{align}
Next, we need to evaluate the matrix element $\mathcal{F}(N_c-1,+;N_c,+)$ which is given by Eqn.~\eqref{mat-el}.  As $B\rightarrow 0$,
\begin{align}\label{Cup+}
\mathcal{C}_{N,+}^\uparrow\approx -\sqrt{\frac{1}{2}+\frac{\Delta}{2\sqrt{2NE_1^2+\Delta^2}}},
\end{align}
and
\begin{align}\label{Cdown+}
\mathcal{C}_{N,+}^\downarrow\approx \sqrt{\frac{1}{2}-\frac{\Delta}{2\sqrt{2NE_1^2+\Delta^2}}}.
\end{align}
Substituting these into Eqn.~\eqref{mat-el} for $\mathcal{F}(N_c-1,+;N_c,+)$, we obtain
\begin{align}
\mathcal{F}(N_c-1,+;N_c,+)\approx\left[-\frac{1}{2}\sqrt{\frac{2E_1^2N_c}{2E_1^2N_c+\Delta^2}}-\frac{E_0\sqrt{N_c}}{\sqrt{2}E_1}\right]^2.
\end{align}
Using the relation between $N_c$ and $\mu$:
\begin{align}
\mu=E_0N_c+\sqrt{2N_cE_1^2+\Delta^2},
\end{align}
we can write
\begin{align}\label{Nc}
N_c=\frac{\mathcal{C}(\mu)}{E_0},
\end{align}
where
\begin{align}\label{C}
\mathcal{C}(\mu)=\mu+mv_F^2-\sqrt{(\mu+mv_F^2)^2+\Delta^2-\mu^2}.
\end{align}
The optical spectral weight of interest (which is simply the Drude weight $W_D$ for $B\rightarrow 0$) is then
\begin{align}
W_D=\frac{e^2}{\hbar}\frac{E_1^2}{8}\left[\sqrt{\frac{2mv_F^2\mathcal{C}(\mu)}{\Delta^2+2mv_F^2\mathcal{C}(\mu)}}+\sqrt{\frac{2\mathcal{C}(\mu)}{mv_F^2}}\right]^2\left.\left(\frac{\partial\mathcal{E}_{N,+}}{\partial N}\right)^{-1}\right|_{N_c}.
\end{align}
But,
\begin{align}
\left.\frac{\partial\mathcal{E}_{N,+}}{\partial N}\right|_{N_c}=E_0\left[1+\frac{mv_F^2}{\sqrt{\Delta^2+2mv_F^2\mathcal{C}(\mu)}}\right]
\end{align}
and hence
\begin{align}
W_D=\frac{e^2}{4\hbar}\mathcal{C}(\mu)\left[1+\frac{mv_F^2}{\sqrt{\Delta^2+2mv_F^2\mathcal{C}(\mu)}}\right],
\end{align}
with $\mathcal{C}(\omega)$ given by Eqn.~\eqref{C}.  Analogous algebra applies to the case of negative $\mu$.  The final formula for the Drude weight which applies to $\mu \lessgtr 0$ is
\begin{align}\label{WD}
W_D=\frac{e^2}{4\hbar}\mathcal{C}(\mu)\left|1+{\rm sgn}(\mu)\frac{mv_F^2}{\sqrt{\Delta^2+2mv_F^2\mathcal{C}(\mu)}}\right|.
\end{align}

\begin{figure}[h!]
\begin{center}
\includegraphics[width=1.0\linewidth]{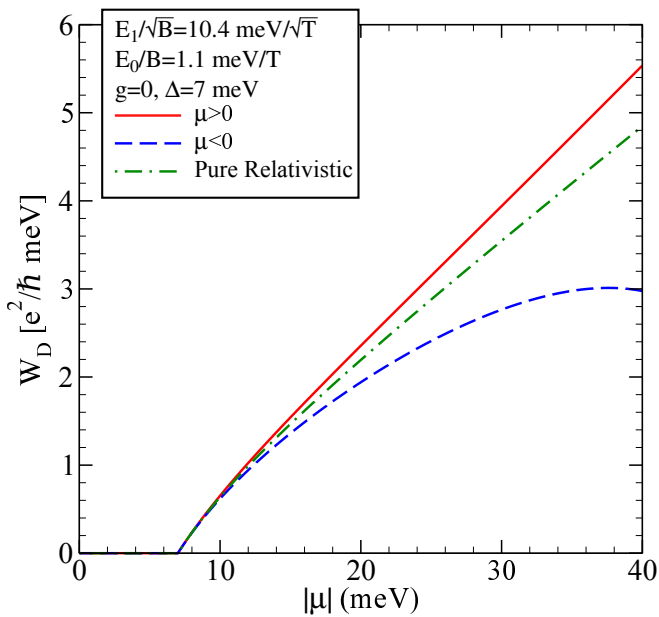}
\end{center}
\caption{\label{fig:Cyc}(Color online) Spectral weight of the Drude peak as a function of $|\mu|$ for positive and negative chemical potential.  The results of a gapped TI are compared to that of a gapped Dirac system.  Particle-hole asymmetry is observed. 
}
\end{figure}

Figure~\ref{fig:Cyc} shows results for the Drude weight $W_D$ of Eqn.~\eqref{WD} when $g=0$ and $\Delta=7$ meV as a function of $|\mu|$ for $\mu>0$ (solid red), $\mu<0$ (dashed blue) and the pure Dirac case ($E_0=0$) (dash-dotted green) for comparison.
It is clear that, while the non-relativistic term is subdominant to the Dirac contribution, it nevertheless makes a significant contribution to the Drude weight as the magnitude of the chemical potential is increased.  The deviations from the dash-dotted green curve are downward for $\mu<0$ and upward for $\mu>0$.  Finally, we note that, for $\Delta=0$, all curves would start at $\mu=0$ rather than $|\mu|=\Delta=7$ meV for the case presented here.  We verified that Eqn.~\eqref{WD} reduces to the known result\cite{ZLi:2015} when $\Delta=0$.  In this case, Eqn.~\eqref{C} can be rewritten in the simpler form
\begin{align}
\mathcal{C}(\omega)=\frac{mv_F^2}{2}\left(1-\sqrt{1+\frac{2\mu }{mv_F^2}}\right)^2.
\end{align}
Using this, the Drude weight for $\Delta=0$ is
\begin{align}
W_D=\frac{e^2}{8\hbar}mv_F^2\left|\left(1-\sqrt{1+\frac{2\mu }{mv_F^2}}\right)\sqrt{1+\frac{2\mu }{mv_F^2}}\right|
\end{align}
which agrees with known result.  Another limit we have checked is the gapped relativistic system ($m\rightarrow\infty$).  After some straight-forward algebra, we obtain
\begin{align}
W_D=\frac{e^2}{8\hbar}\frac{\mu^2-\Delta^2}{|\mu|}.
\end{align}

\begin{figure}[h!]
\begin{center}
\includegraphics[width=1.0\linewidth]{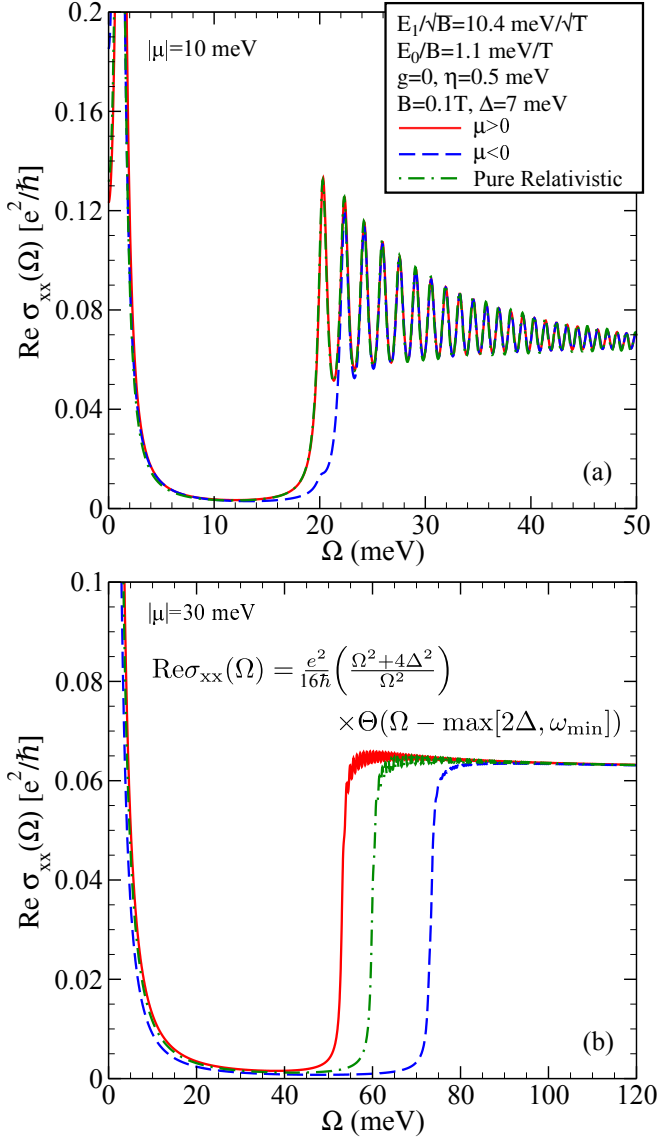}
\end{center}
\caption{\label{fig:Cond-Background}(Color online) Longitudinal magneto-optical conductivity for small $B$ and large $|\mu|$.  The usual $B=0$ interband background becomes resolved as $\mu$ increases [frame (b)].  The results for a gapped TI and $\mu \lessgtr 0$ are compared to a gapped Dirac system.  Particle-hole asymmetry is evident.
}
\end{figure}

When discussing the $B\rightarrow 0$ limit, it is of interest to examine how the interband contribution evolves into an absorption background.  Numerical results obtained from Eqn.~\eqref{sigmaxx} for Re$\sigma_{xx}(\Omega)$ are presented in Fig.~\ref{fig:Cond-Background}.
Here, we use $g=0$, $\eta=0.5$ meV, $B=0.1$T and $\Delta=7$ meV. In frame (a), $|\mu|=10$ meV and we show both positive (solid red) and negative (dashed blue) $\mu$. Again, we include the pure relativistic result for comparison (dash-dotted green).  From this we note that the Schr\"odinger term has a negligible effect on the background for positive $\mu$.  Later, we will show that this is expected for small $\mu$.  In Fig.~\ref{fig:Cyc}, we saw a similar result for the Drude weight, where deviations become strong only for larger values of $\mu$.  Returning to Fig.~\ref{fig:Cond-Background}, we observe that, as $\Omega$ is reduced towards $20$ meV (i.e. $2|\mu|$), the background (defined as the envelope through the center of oscillations) increases above the universal value [$e^2/(16\hbar)$] seen at higher $\Omega$.  In addition, for a TI, we do not have particle-hole symmetric responses.  The onset of the background for $\mu>0$ is lower in energy that that of $\mu<0$.  This difference in onset frequency is clearly seen in frame (b) where we use $|\mu|=30$ meV.  In this case, the solid red curve has its interband edge at an energy considerably below that of the pure relativistic system (dash-dotted green).  Conversely, the $\mu<0$ onset has been pushed upwards by an even greater amount.  

\begin{figure}[h!]
\begin{center}
\includegraphics[width=1.0\linewidth]{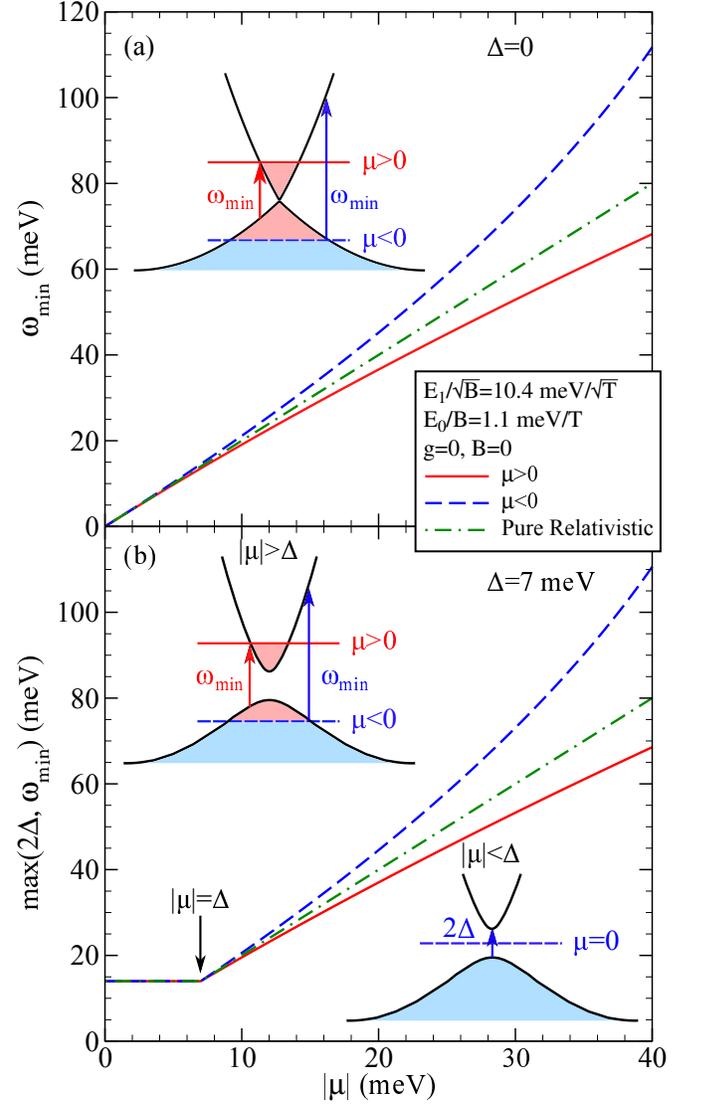}
\end{center}
\caption{\label{fig:wmin}(Color online) Minimum energy required for interband transitions in a (a) gapless and (b) gapped TI as a function of $|\mu|$.  The results for $\mu \lessgtr 0$ are compared to the pure Dirac system.  Particle-hole asymmetry is observed.  Insets: schematic plots of the low-energy band structure for $\pm|\mu|$.  The lowest energy interband transition is marked by the blue ($\mu\leq 0$) or red ($\mu>0$) arrow.
}
\end{figure}

The onset (and its variation with $\mu$, $\Delta$ and $m$) can be calculated with the help of the illustrative band structures inserted in Fig.~\ref{fig:wmin}.
The interband transition with minimum photon energy $\omega_{\rm min}$ is shown as a red arrow for $\mu>0$ and the longer blue arrow corresponds to $\mu<0$.  For positive chemical potential, the momentum $k_c$ associated with this transition is given by $\mu=E_+(k_c)$ while for $\mu<0$ it is $\mu=E_-(k_c)$.  In both cases, $\omega_{\rm min}=E_+(k_c)-E_-(k_c)$.  After much algebra, we find
\begin{align}\label{wmin}
\omega_{\rm min}=2\sqrt{\Delta^2+2(mv_F^2)^2\left[1+\frac{\mu }{mv_F^2}-\sqrt{1+\frac{2\mu }{mv_F^2}+\frac{\Delta^2}{(mv_F^2)^2}}\right]}
\end{align}
which is valid for $\mu\lessgtr 0$.  When $\Delta=0$, this reduces correctly to the known result\cite{ZLi:2015}
\begin{align}
\omega_{\rm min}=2mv_F^2\left|1-\sqrt{1+\frac{2\mu }{mv_F^2}}\right|
\end{align}
and to $2|\mu|$ when $m\rightarrow\infty$\cite{Gusynin:2006a,Gusynin:2009,Schmeltzer:2013,Peres:2013} (i.e. the Schr\"odinger contribution is neglected).  Finally, we note that the interband transitions are also bounded above.  As long as the non-relativistic contribution is much smaller than the relativistic part of the Hamiltonian, an upper cutoff will come from cutting off at the Brillouin zone boundary where the low-energy Hamiltonian ceases to be valid.

We now derive a simple analytic expression for the interband background in the limit of $B\rightarrow 0$.  More precisely, we assume that the magnitude of the chemical potential $|\mu|$ is much greater than $E_1$ and $E_0$.  This is the same limit used in the previous discussion of $W_D$.  We start with our general expression for Re$\sigma_{xx}(\Omega)$ at finite $B$ [Eqn.~\eqref{sigmaxx}] and take the appropriate limit.  For simplicity, we assume $\mu>0$ ($\mu<0$ is analogous).  The two interband transitions of interest for a given $N^s$ are $N^-\rightarrow(N+1)^+$ and $(N+1)^-\rightarrow N^+$.  For $B\rightarrow 0$, the $E_0$ term in the energy transition drops out and both transitions have the same energy $2\sqrt{2NE_1^2+\Delta^2}$.  Taking $\eta\rightarrow 0$ in the Lorentzians, we arrive at
\begin{align}\label{sigxx-B0}
{\rm Re}\sigma_{xx}(\Omega)&=\frac{e^2}{\hbar}\frac{E_1^2}{2}\sum_N\frac{1}{\Omega}\delta(\Omega-2\sqrt{2NE_1^2+\Delta^2})\notag\\
&\times[\mathcal{F}(N,+;N+1,-)+\mathcal{F}(N-1,-;N,+)],
\end{align}
where we have assumed the final state $(N+1)^+$ or $N^+$ to be unoccupied while the initial state of the transition was occupied.  This allows us to replace the thermal factors by 1 or 0.  In doing so, a critical value of $N$ is defined with $\mu=\mathcal{E}_+(N_c)$.  This defines a minimum energy for interband transitions (namely $\Omega=2\sqrt{2N_cE_1^2+\Delta^2}$).  As a result of the $\delta$-function in Eqn.~\eqref{sigxx-B0}, we need photon energies $\Omega\geq 2\sqrt{2N_cE_1^2+\Delta^2}$ to get a nonzero result.  For $\mu<\Delta$, it is clear from the inset of Fig.~\ref{fig:wmin}(b) that $N_c=0$ and we need $\Omega>2|\Delta|$.  Conversely, for $\mu>\Delta$, $N_c$ is not zero but is instead given by Eqn.~\eqref{Nc}; hence, Re$\sigma_{xx}(\Omega)$ will be zero until the photon energy $\Omega$ is $\geq 2\sqrt{2N_cE_1^2+\Delta^2}=2\sqrt{2mv_F^2\mathcal{C}(\mu)+\Delta^2}$ which reduces to our previous result for $\omega_{\rm min}$ [Eqn.~\eqref{wmin}] obtained from momentum space considerations.  Here it is derived from the Kubo formula with LLs in the limit $B\rightarrow 0$.  For $|\mu|\gg E_1$ and $E_0$, $N_c$ is large and consequently, Eqn.~\eqref{sigxx-B0} reduces to
\begin{align}\label{sigxx-B0-BigN}
{\rm Re}\sigma_{xx}(\Omega)&=\frac{e^2}{\hbar}\frac{E_1^2}{2\Omega}\sum_N\delta(\Omega-2\sqrt{2NE_1^2+\Delta^2})\notag\\
&\times[\mathcal{F}(N,+;N,-)+\mathcal{F}(N,-;N,+)]\notag\\
&\times\Theta(\Omega-{\rm max}[2\Delta,\omega_{\rm min}]).
\end{align} 
In the limit of interest,
\begin{align}
\mathcal{C}_{N-}^\uparrow=\sqrt{\frac{1}{2}-\frac{\Delta}{2\sqrt{2NE_1^2+\Delta^2}}}
\end{align}
and
\begin{align}
\mathcal{C}_{N-}^\downarrow=\sqrt{\frac{1}{2}+\frac{\Delta}{2\sqrt{2NE_1^2+\Delta^2}}}.
\end{align}
Using Eqns.~\eqref{Cup+} and ~\eqref{Cdown+}, we see that the combination
\begin{align*}
\sqrt{N}\left(\mathcal{C}_{N-}^\uparrow\mathcal{C}_{N+}^\uparrow+\mathcal{C}_{N-}^\downarrow\mathcal{C}_{N+}^\downarrow\right)
\end{align*}
which enters the optical matrix elements is zero leaving us with $\mathcal{F}(Ns;Ms^\prime)=(\mathcal{C}_{Ms^\prime}^\uparrow \mathcal{C}_{Ns}^\downarrow)^2$. Therefore, the sum of $\mathcal{F}$'s which enters Eqn.~\eqref{sigxx-B0-BigN} becomes $(\Omega^2+4\Delta^2)/(2\Omega^2)$.  Changing the sum over $N$ into an integral, the $\delta$-function contributes a factor of $\Omega/(4E_1^2)$ so that
\begin{align}\label{xx-B0-lim}
{\rm Re}\sigma_{xx}(\Omega)&=\frac{e^2}{16\hbar}\frac{\Omega^2+4\Delta^2}{\Omega^2}\Theta(\Omega-{\rm max}[2|\Delta|,\omega_{\rm min}]).
\end{align} 
This correctly reduces to the known result that the background conductivity is $e^2/(16\hbar)$ (excluding spin and valley degeneracies) when $\Delta=0$, independent of the value of $m$\cite{ZLi:2015}.  In the limit of finite $\Delta$ and $m\rightarrow\infty$, we get another known result\cite{Gusynin:2009}
\begin{align}
{\rm Re}\sigma_{xx}(\Omega)&=\frac{e^2}{16\hbar}\frac{\Omega^2+4\Delta^2}{\Omega^2}\Theta(\Omega-{\rm max}[2|\Delta|,2|\mu|]),
\end{align}
which differs from Eqn.~\eqref{xx-B0-lim} in that $2|\mu|$ replaces $\omega_{\rm min}$.  Note when $|\mu|<|\Delta|$, the interband background is twice its value for $\Omega\gg\Delta$.  In Fig.~\ref{fig:wmin}(a), we show our results for $\omega_{\rm min}(\mu)$ when $\Delta=0$.  The red curve applies to $\mu>0$ and the dashed blue to $\mu<0$.  Both curves deviate from the dash-dotted green case which applies to the pure relativistic limit when $m\rightarrow\infty$.  Frame (b) shows similar results for $\Delta=7$ meV where max$[2|\Delta|,2|\mu|]$ is plotted.  As expected, for large $|\mu|$ compared to $|\Delta|$, the results do not significantly depend on the gap value.  These results show that the onset of interband transitions is changed by the introduction of a subdominant non-relativistic term in our Hamiltonian; while, there is no change in the height of the interband background even when a gap is present.  We also see significant particle-hole asymmetry in the interband onset for large $|\mu|$.

\section{Discussion and Conclusion}

We have calculated the magneto-optical conductivity of a topological insulator with particular attention paid to the interplay between a gap ($\Delta$) and the subdominant non-relativistic Schr\"odinger mass ($m$).  When the gap is zero, the non-relativistic quadratic-in-momentum piece of the Hamiltonian is known to split the interband optical absorption lines into doublets because the energy of the $N^s=N^-$ to $(N+1)^+$ transition is not degenerate with $(N-1)^-$ to $N^+$ as they would be in the pure Dirac system.  The magnitude of the splitting is given by the Schr\"odinger magnetic scale $E_0=\hbar eB/m$.  For $\Delta=0$, the optical spectral weight under each doublet line is nearly equal.  This is changed when a finite gap is included.  The change in spectral weight depends on the sign of the gap.  For $\Delta>0$, the lower energy line loses much of its intensity while the upper line gains spectral weight.  For $\Delta<0$, the opposite occurs.  Differences are also present in the real part of the dynamical Hall conductivity for which the structures associated with interband transitions show peaks with broadened tops.  This is distinctly different from the pure relativistic material.  An additional effect of introducing a gap and Schr\"odinger mass is the shift in position of the absorption lines in Re$\sigma_{xx}(\Omega)$ and the corresponding structures in Re$\sigma_{xy}(\Omega)$.  

The mass term is known to drop out completely from the height of the universal interband background when $\Delta=0$ and the magnetic field $B$ is zero.  When a gap is included in the pure relativistic system, it is known to introduce a modulating factor $(\Omega^2+4\Delta^2)/\Omega^2$ to the background.  Here, we derive an analytic expression for the interband transitions when $B=0$ and $\Delta$ and $m$ are both finite.  We proceed directly from our general expression for the finite $B$ conductivity and formally take $B\rightarrow 0$.  We find that the Schr\"odinger mass drops out from the background and the modulating factor due to the gap remains unchanged.  Consequently, only the onset frequency of interband transitions is affected by $m$.  This onset depends on $\Delta$, $m$ and $\mu$.  It does not depend on the sign of the gap; however, the sign of the chemical potential is integral making it particle-hole asymmetric.

A similar situation is found for the real part of the Hall conductivity.  In general, it is modified through the introduction of a mass term or a gap.  However, the dc limit which gives the quantized Hall plateaus is not altered.  Of course, both the gap and mass change the value of chemical potential at which a transition to a new plateau occurs.  This transition is also dependent on the sign of the gap.  At charge neutrality, the Hall quantization (in units of $e^2/h$) is $-1/2$ for $\Delta<0$; while, for $\Delta>0$ is is $1/2$.  This can be traced to the $N=0$ level which sits at $(1+g)E_0/2-\Delta$ (where $g$ is the Zeeman splitting) which is negative for $\Delta>0$ and $|\Delta|>(1+g)E_0/2$, but is positive for $\Delta<0$.  For $\Delta>0$, the magnetic field can be increased sufficiently to push the $N=0$ Landau level from negative to positive energy.  This leads to a switch in sign of the $\mu=0$ Hall conductivity.  The critical value of $B$ is $B_c=(2\Delta m)/([1+g]\hbar e)$.  At this value of $B$, the charge neutral intraband absorption line in the longitudinal conductivity shows a jump in spectral weight and onset energy for $\Delta>0$.  When $\Delta<0$, no such singular behaviour is found.

\begin{acknowledgments}
This work has been supported by the Natural Sciences and Engineering Research Council of Canada and, in part, by the Canadian Institute for Advanced Research.
\end{acknowledgments}

\bibliographystyle{apsrev4-1}
\bibliography{TI-CON}

\end{document}